\documentclass[aps,prl,twocolumn,showpacs,preprintnumbers,amsmath,amssymb]{revtex4-1}

\usepackage{upgreek}
\usepackage{graphicx} 
\usepackage{bm} 

\renewcommand{\vec}[1]{\bm{\mathbf{#1}}}
\newcommand{\hvec}[1]{\hat{\vec{#1}}\vphantom{#1}}
\newcommand{\mtrx}[1]{\boldsymbol{#1}}

\DeclareMathOperator{\const}{const}

\newcommand{\He}[1]{\textsuperscript{#1}He}
\newcommand{\micro}{\ensuremath{\upmu}}
\newcommand{\um}{\micro{}m}

\newcommand{\Dperp}{\Delta_\perp}
\newcommand{\Dpar}{\Delta_\parallel}

\newcommand{\aver}[1]{\overline{#1}\rule{0pt}{0.75em}}

\usepackage{color}

\begin{document}

\title{Evidence for a Spatially-Modulated Superfluid Phase of \He3 under Confinement}

\author{Lev V. Levitin}
\email{l.v.levitin@rhul.ac.uk}
\affiliation{Department of Physics, Royal Holloway University of London, Egham, Surrey, TW20 0EX, UK}
\author{Ben Yager}
\altaffiliation[Now at ]{Institute for Quantum Computing, University of Waterloo, Waterloo, Ontario N2L 3G1, Canada}
\affiliation{Department of Physics, Royal Holloway University of London, Egham, Surrey, TW20 0EX, UK}
\author{Laura Sumner}
\altaffiliation[Now at ]{Imperial College London, London SW7 2AZ, UK}
\affiliation{Department of Physics, Royal Holloway University of London, Egham, Surrey, TW20 0EX, UK}
\author{Nikolay Zhelev}
\altaffiliation[Now at ]{Corning Incorporated, USA}
\affiliation{Department of Physics, Cornell University, Ithaca, NY, 14853 USA}
\author{Robert G. Bennett}
\altaffiliation[Now at ]{Vantage Power Ltd, London, UB6 0FD, UK}
\affiliation{Department of Physics, Cornell University, Ithaca, NY, 14853 USA}
\author{Jeevak M. Parpia}
\affiliation{Department of Physics, Cornell University, Ithaca, NY, 14853 USA}
\author{Brian Cowan}
\affiliation{Department of Physics, Royal Holloway University of London, Egham, Surrey, TW20 0EX, UK}
\author{Andrew J. Casey}
\affiliation{Department of Physics, Royal Holloway University of London, Egham, Surrey, TW20 0EX, UK}
\author{John Saunders}
\affiliation{Department of Physics, Royal Holloway University of London, Egham, Surrey, TW20 0EX, UK}

\date{15 January 2019}

\begin{abstract}
In superfluid \He3-B confined in a slab geometry,
domain walls between regions of different order parameter orientation
are predicted to be energetically stable.
Formation of the spatially-modulated superfluid \emph{stripe phase} has been proposed.
We confined \He3 in a 1.1\,\um{} high microfluidic cavity
and cooled it into the B phase at low pressure,
where the stripe phase is predicted.
We measured the surface-induced order parameter distortion with NMR,
sensitive to the formation of domains.
The results rule out the stripe phase, but are consistent with 2D modulated superfluid order.
\end{abstract}

\maketitle

The pairing of fermions to form a superfluid or superconductor at sufficiently
low temperatures is a relatively ubiquitous phenomenon~\cite{bib:BCS, bib:BCS50}.
Examples include: electrically conducting systems from metals
to organic materials to metallic oxides \cite{bib:SC:review};
neutral atoms from \He3~\cite{bib:Leggett:RMP, bib:VW}
to ultracold fermionic gases~\cite{bib:cold:gases};
and astrophysical objects such as neutron stars and pulsars~\cite{bib:Migdal:nuclearSC}.
In the most straightforward case the pairs form
a macroscopic quantum condensate which is spatially uniform.
In type-II superconductors a spatially inhomogeneous state, the Abrikosov flux lattice,
arises in a magnetic field~\cite{bib:Abrikosov:flux}.
Its origin is the negative surface energy between normal and superconducting regions.
However the realisation and experimental identification
of states with spatially-modulated superfluid/superconducting order has proved challenging.

The Fulde-Ferrell-Larkin-Ovchinnikov (FFLO) state \cite{bib:LO,bib:FF},
has been predicted to arise in \emph{spin-singlet} superconductors.
An imbalance between spin-up and spin-down Fermi momenta,
driven by ferromagnetic interactions or high magnetic fields,
induces pairing with non-zero centre of mass momentum.
This results in both the order parameter and
the spin density oscillating in space with the same wavevector.
The FFLO state is predicted to intervene beyond the Pauli limiting field,
inhibiting the destruction of superconductivity~\cite{bib:MSh:FFLO:review}.
It requires orbital effects to be weak, restricting possible materials for its observation.
There is evidence of the FFLO state in the layered organic
superconductors $\kappa$-(BEDT-TTF)$_2$Cu(NCS)$_2$ and
$\beta''$-(ET)$_2$SF$_5$CH$_2$CF$_2$SO$_3$~\cite{bib:Wosnitza:organic:FFLO,
bib:Mayaffre:organic:nmr, bib:Agosta:organic:calorimetry,
bib:Beyer:organic:calorimetry,bib:Koutroulakis:NMR}, and in the canonical
heavy fermion superconductor CeCu$_2$Si$_2$~\cite{bib:Kitagawa:NMR:CeCu2Si2}.
Previously identified as FFLO~\cite{bib:Bianchi:CeCoIn5:FFLO},
a more complex state, with intertwined p-wave pair density wave (PDW)
and spin density wave has been proposed in the heavy fermion d-wave superconductor
CeCoIn$_5$~\cite{bib:Koutroulakis:CeCoIn5:NMR,
bib:Duk:CeCoIn5:intertwined,bib:Duk:CeCoIn5:reskappa}.
Elsewhere a PDW commensurate with a charge density wave,
has been clearly demonstrated in the d-wave cuprate superconductor
Bi$_2$Sr$_2$CaCu$_2$O$_{8+x}$~\cite{bib:Hamidian:PDW}.
In the ultracold fermionic gas $^6$Li, superfluidity with imbalanced spin populations
has been observed~\cite{bib:Zwierlein:imbalanced:sf}
with thermodynamic evidence consistent with FFLO~\cite{bib:Revelle:imbalanced:crossover}.
In addition to these condensed matter systems
it has been proposed that quantum chromodynamics may provide
a pathway to inhomogeneous superconductivity, potentially realised
in astrophysical objects~\cite{bib:Casalbuoni:LOFF:review}. 

In general the order parameter modulation is expected to be more complex than
the model FFLO state~\cite{bib:MSh:FFLO:review,bib:Dutta:FFLO:protocol}.
Potential examples are: 1D domain walls of thickness much smaller
than the width of domains; 2D modulated structures,
involving multiple wavevectors~\cite{bib:MSh:FFLO:review}.
Furthermore, nucleation barriers and metastability may inhibit
the formation of periodic states~\cite{bib:Dutta:FFLO:protocol}.

In this paper we report experimental investigation of
a predicted spatially modulated state in the topological p-wave,
\emph{spin-triplet}, superfluid $^3$He~\cite{bib:VorontsovSauls:stripes}.
This requires the superfluid to be confined in a thin cavity of uniform thickness.
At the heart of this predicted \emph{stripe phase} is the stabilisation
of a hard domain wall, of thickness comparable to the superfluid coherence length,
in superfluid $^3$He-B under confinement in a slab geometry.
These B-B domain walls were first classified in~Ref.~\cite{bib:SalomaaVolovik:cosmological},
and the analogy drawn with cosmic domain walls.
Their stability in the bulk, and possible evidence for their observation
is discussed in~\cite{bib:Thuneberg:hardwalls}.
Under confinement the presence of the domain wall reduces surface pair-breaking,
and can result in a negative domain wall energy, leading to the formation of the stripe
phase~\cite{bib:VorontsovSauls:stripes, bib:WimanSauls:strongstripes, bib:Aoyama:stripes}.

In superfluid $^3$He the nuclear spins constitute the spin part of the pair wavefunction;
thus nuclear magnetic resonance (NMR) is widely used to provide a direct fingerprint
of the superfluid order parameter~\cite{bib:Leggett:RMP,bib:VW}. NMR has been predicted
to distinguish clearly between the striped and translationally-invariant states
of the B phase~\cite{bib:us:SlabPRL,bib:WimanSauls:strongstripes}.

To optimise the formation of stripes in this work we chose
a slab geometry of height $D=1.1$\,\um{}, where
the B phase is stable down to zero pressure~\cite{bib:Cornell:SlabTO}.
The stripe phase was originally predicted in the weak-coupling limit
of Bardeen-Cooper-Schrieffer theory~\cite{bib:VorontsovSauls:stripes},
while the strong-coupling corrections to this theory in general favour the A phase
and suppress the stability of stripes~\cite{bib:WimanSauls:strongstripes}.
At present the strong coupling effects are not fully understood theoretically,
leaving the stability of the stripe phase an open
question~\cite{bib:Cornell:SlabTO, bib:WimanSauls:strongstripes}.
We performed the experiment at low pressure to minimise the strong-coupling effects.

Here we show that under our experimental conditions the stripe phase is clearly ruled out.
However, there is NMR evidence for a spatially modulated superfluid
of two-dimensional morphology, similar to states discussed in the context of
FFLO~\cite{bib:MSh:FFLO:review}; we term this \emph{polka-dot}.

The 3$\times3$ matrix order parameter of a p-wave superfluid $^3$He
allows for multiple superfluid phases with different broken symmetries
and topological invariants~\cite{bib:Leggett:RMP,bib:VW}.
In the bulk, at low pressure and magnetic field the stable state
is the quasi-isotropic B phase with order parameter matrix
$\mtrx A =  e^{i\phi} \mtrx R \Delta$, where $\Delta$ is the energy gap,
isotropic in the momentum space, $\phi$ is the superfluid phase
and $\mtrx R = \mtrx R (\hvec n, \theta)$ is the matrix of
relative spin-orbit rotation, parametrised by angle $\theta$ and axis $\hvec n$.
This quasi-isotropic phase is relatively easily distorted by magnetic field or flow.
Under confinement the distortion is strong and spatially inhomogeneous,
induced by surface pair-breaking.
In a slab normal to the $z$-axis, the order parameter is predicted to take the form
\begin{equation}\label{eq:Bplanar}
\mtrx A(z) = e^{i\phi} \mtrx R \begin{pmatrix}
\Dpar(z) & &\\ & \!\!\Dpar(z)\!\! & \\ & & \Dperp(z)\end{pmatrix},
\end{equation}
with $0 \le \Dperp < \Dpar$ due to stronger surface pair-breaking
of Cooper pairs with orbital momentum parallel to the slab surface~\cite{bib:Nagai:AB}.
This distortion is named \emph{planar} after the planar phase,
in which $\Dperp = 0$~\cite{bib:VW}.

\begin{figure}[t!]
\centerline{\includegraphics{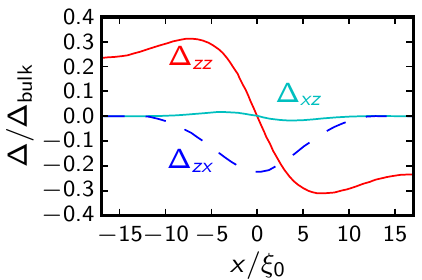}}
\caption{Domain wall at the heart of the stripe phase in \He3-B~\cite{bib:VorontsovSauls:stripes}.
Shown here are the elements of the energy gap matrix $\mtrx\Delta$,
such that $\mtrx A = e^{i\phi} \mtrx R \mtrx\Delta$.
When the domain wall is absent $\mtrx\Delta$ is diagonal, see Eq.~\eqref{eq:Bplanar}.
Crossing the domain wall lying in $yz$-plane at $x=0$, $\Delta_{zz}$ changes
from $\Dperp$ to $-\Dperp$,
and off-diagonal elements $\Delta_{xz}$ and $\Delta_{zx}$ emerge, while
$\Delta_{xx}$ and $\Delta_{yy}$ (not shown) remain close to $\Dpar$.
The gap amplitudes were calculated $z=2.5\xi_0$ away from one of the surfaces
in a $D=10\xi_0$ thick slab at $T=0.5T_c^{\text{bulk}}$;
$\xi_0$ is the Cooper pair diameter, $\Delta_{\text{bulk}}$ is the bulk B phase gap.}
\label{fig:domainwall}
\end{figure}

The order parameter~\eqref{eq:Bplanar} has a large manifold of orientations,
determined by $\phi$, $\hvec n$ and $\theta$, allowing domain walls between regions
of different orientations~\cite{bib:SalomaaVolovik:cosmological, bib:Thuneberg:hardwalls}.
Domain walls where $\Dperp$ changes sign, shown in Fig.~\ref{fig:domainwall},
are predicted to have negative surface energy in the B phase confined in a thin
slab, close to the A-B transition~\cite{bib:VorontsovSauls:stripes, bib:stripe:sign}.
As a consequence the slab of \He3-B would spontaneously break into domains,
until the domain walls get close enough that their mutual repulsion becomes significant.
This is predicted to result in the periodic stripe phase
with a typical domain size $W$ of order $D$~\cite{bib:VorontsovSauls:stripes,
bib:Aoyama:stripes, bib:WimanSauls:strongstripes}.
Phases with spontaneously broken translational invariance are also predicted to stabilise
in \He3 confined to narrow pores~\cite{bib:Ayoma:pores,bib:WimanSauls:pores}
and in films of d-wave superconductors~\cite{bib:Vorontsov:dwavestripes}.

In this experiment we performed pulsed NMR studies on a slab of \He3 confined
in a $D=1144 \pm 7$\,nm thick silicon-glass microfluidic cavity,
in a field of 31~mT perpendicular to the slab
(corresponding to \He3 Larmor frequency $f_{\text L} = 1.02$\,MHz),
using the setup described in Refs.~\cite{bib:us:SlabScience, bib:supplementary}.
The measurements were performed at low pressure $P=0.03$~bar,
where the bulk superfluid transition temperature $T_c^{\text{bulk}} = 0.93$\,mK,
and close to specular scattering, achieved by preplating the cell walls with
a 64\,\micro{}mol/m$^2$ ($\sim5$ atomic layers) \He4 film.

\begin{figure}[t!]
\includegraphics{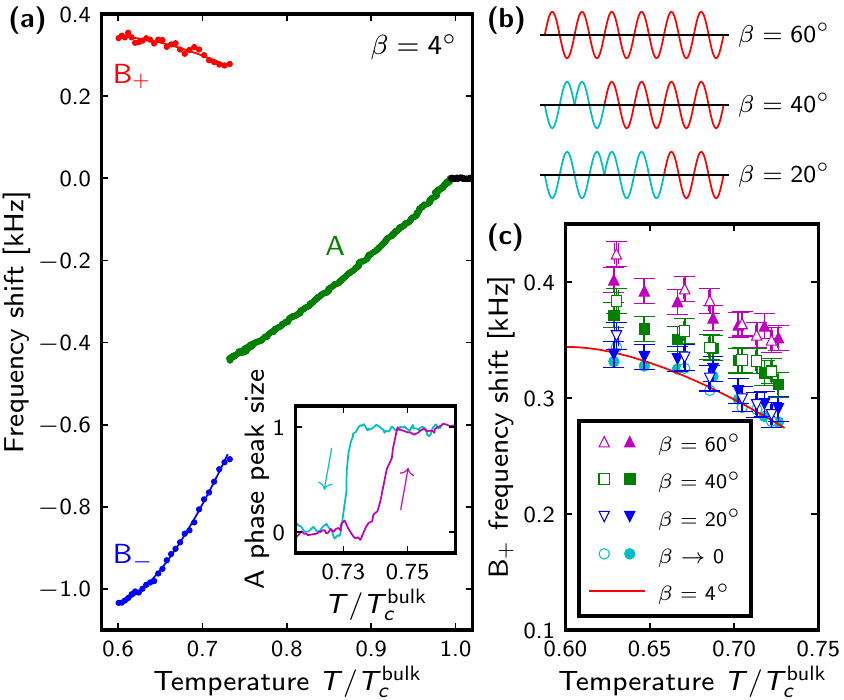}
\caption{NMR measurements on a $D=1.1$\,\um{} slab of superfluid \He3.
(a) Signatures of the A and B phases observed with small tipping pulses.
At the second-order normal-superfluid transition
a negative frequency shift develops as the slab enters the A phase
with the dipole-unlocked orientation.
On further cooling a first-order phase transition
into the B phase with a planar distortion occurs,
where two spin-orbit orientations B$_+$ and B$_-$ are observed.
Inset: sharp A-B transition with small hysteresis.
(b) Technique for applying a set of pulses with different tipping angle $\beta$
but equal heating: all pulses coincide in length and amplitude;
$\beta$ is reduced by applying the initial section
of the pulse with a $180^\circ$ phase shift,
which cancels out a similar section that follows (both light blue),
so only the remainder of the pulse (red) tips the spins.
(c) Initial frequency shifts in B$_+$ after such pulses yield
$\partial\Delta f_+ / \partial \cos\beta$ and $\Delta f_+(\beta\to 0)$.
Good agreement between $\Delta f_+(\beta\to 0)$ obtained here
and $\Delta f_+(4^\circ)$, smoothed from (a),
indicates that the heating due to the 20-60$^\circ$ pulses is negligible.
Open/filled symbols show data taken on step-wise warm-ups
after a fast/slow cool-down from the A phase.}
\label{fig:NMR}
\end{figure}

We first mapped the phase diagram with small tipping angle, $\beta = 4^\circ$,
NMR pulses, Fig.~\ref{fig:NMR}a.
The A-B transition was observed at $T_{\text{AB}} = 0.7 T_c^{\text{bulk}}$
in agreement with torsional oscillator measurements,
with a 1.08\,\um{} cavity \cite{bib:Cornell:SlabTO}.
As we previously observed in the 0.7\,\um{} cavity, the B phase nucleated stochastically
in two spin-orbit orientations with distinct NMR signatures:
stable B$_+$ and metastable B$_-$ \cite{bib:us:SlabPRL,bib:us:SlabScience}.

The magnitudes of the frequency shifts of translationally invariant
B$_+$ and B$_-$, $\Delta f_+$ and $\Delta f_-$,
are determined by averages of the gap structure across the cavity \cite{bib:us:SlabPRL}:
$\langle \Dpar^2\rangle$, $\langle \Dpar \Dperp \rangle$ and $\langle \Dperp^2 \rangle$.
In case of the putative spatially-modulated phase the averaging is also
performed in the plane of the slab.
This procedure is valid when the width of the stripes $W$ is smaller
than the dipole length $\xi_{\text D} \approx 10$\,\um{} \cite{bib:us:SlabPRL},
a condition predicted to hold for this cavity ($W \approx D\sqrt{3} \ll \xi_{\text D}$),
except very close to the stripe-to-B transition \cite{bib:WimanSauls:strongstripes}.
In the stripe phase $\langle \Dpar \Dperp \rangle = 0$
due to $\Dperp$ having opposite sign in the adjacent domains \cite{bib:us:SlabPRL}.
This has clear signatures in the NMR response, as a function of tipping angle $\beta$.

We define the dimensionless gap distortion parameters
\begin{equation}\label{eq:qQ}
\aver q = \langle\Dpar\Dperp\rangle / \langle\Dpar^2\rangle,
\qquad
\aver Q = \sqrt{\langle \Dperp^2\rangle / \langle \Dpar^2\rangle}.
\end{equation}
The tipping angle dependence of the frequency shift of B$_+$, below the so called ``magic angle'', $\beta^* > 104^\circ$,
scaled by the small-tipping-angle shift of B$_-$ is given by \cite{bib:us:SlabPRL}
\begin{subequations}\label{eq:fshiftqQ}
\begin{eqnarray}
\dfrac{\Delta f_+(\beta\to0)}{\Delta f_-(\beta\to 0)} &=& \dfrac{2\aver Q^2 - \aver q^2 - 1}{1 + 2\aver Q^2},\\
\dfrac{\partial\Delta f_+(\beta) / \partial \cos\beta}{\Delta f_-(\beta\to 0)} &=& \dfrac{2\aver Q^2 - 2\aver q^2}{1 + 2\aver Q^2},
\end{eqnarray}
\end{subequations}
where $\beta$ is the tip angle.
We further note that: $\Delta f_-(\beta)$ does not depend on $\aver q$;
the magic angle at which there is a kink in $\Delta f_+(\beta)$
is given by $\beta^* = \arccos (\aver q - 2)/(2\aver q + 2)$ \cite{bib:us:SlabPRL}.
Thus there is no magic angle expected for the stripe phase, for which $\aver q=0$. 

\begin{figure}[t!]
\includegraphics{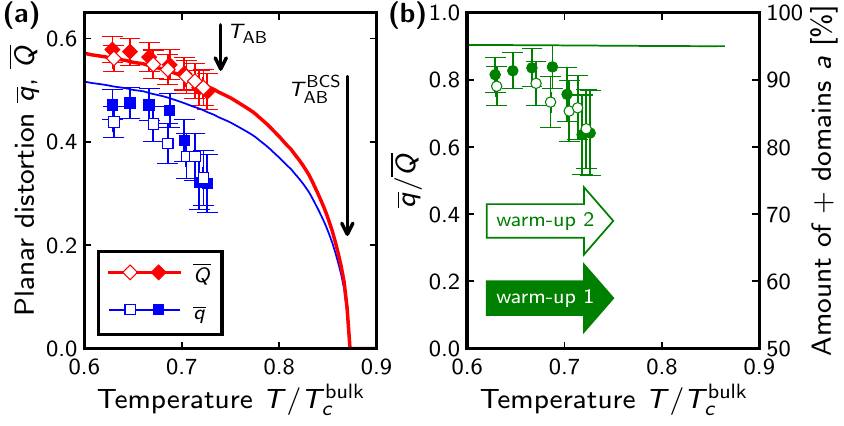}
\caption[NMR measurement of the planar distortion]{
Temperature dependence of planar distortion parameters $\aver q$, $\aver Q$ (a)
and their ratio (b) inferred from NMR measurements, Fig.~\ref{fig:NMR}.
Solid lines show weak-coupling calculations for
translationally-invariant B phase \cite{bib:VorontsovSauls:Aphase}.
In the absence of strong coupling the temperature of the AB transition
$T_{\text{AB}}^{\text{BCS}}$ is higher than $T_{\text{AB}}$ observed experimentally.
While $\aver Q$ is in agreement with the theory,
$\aver q$ gets progressively reduced approaching $T_{\text{AB}}$.
We interpret this in terms of development of domains with opposite
sign of $\Dperp$ on warming.
The right-hand vertical axis in (b) estimates the fraction of the slab occupied
by the majority domains, taken here to have positive $\Dperp$ \cite{bib:footnote:Dperp},
under a qualitative assumption of step-like energy gap profile,
Eq.~\eqref{eq:crude:fraction}. Any systematic difference between
measurements taken while warming after a fast and a slow cool-down through
A-B transition (open/filled symbols) is small.}
\label{fig:qQ}
\end{figure}

Application of large tipping pulses in our setup results in rapid heating of
the confined helium via an unidentified mechanism, that previously restricted
measurements of the planar distortion to temperatures well below $T_{\text{AB}}$
\cite{bib:us:SlabPRL,bib:supplementary}.
Here we focus on moderate pulses, $\beta \lesssim 60^\circ$, that allow us
to probe the temperature dependence of the distortion parameters up to $T_{\text{AB}}$,
according to Eq.~\eqref{eq:fshiftqQ}.
In order to measure the tipping-angle dependence of the frequency shift at
constant temperature, we developed a scheme for applying pulses with different
$\beta$ while inducing identical heating, shown in Fig.~\ref{fig:NMR}b.
Triplets of such pulses with $\beta = 20^\circ$ to $60^\circ$ were applied to B$_+$
during step-wise warm-ups after a fast (at approx.~40\,\micro{}K/min rate)
and slow (4\,\micro{}K/min) cool-down through the A-B transition.
We inferred  $\partial\Delta f_+ / \partial \cos\beta$ and $\Delta f_+(\beta\to0)$
from the data shown in Fig.~\ref{fig:NMR}c and
confirmed the heating effects to be negligible for the chosen pulses.
Combining with $\Delta f_-(4^\circ)$ at the same temperature,
Fig.~\ref{fig:NMR}a, we determine the planar distortion parameters
$\aver q$ and $\aver Q$ through Eq.~\eqref{eq:fshiftqQ}, shown in Fig.~\ref{fig:qQ}.

In the above analysis the off-diagonal elements of the gap matrix near domain walls
(see Fig.~\ref{fig:domainwall}) have been neglected.
Incorporating the detailed gap structure at the domain walls into the NMR model
leaves the signature of the stripe phase, $\aver q = 0, \aver Q > 0$,
virtually unchanged \cite{bib:WimanSauls:strongstripes}.
We can therefore conclude unambiguously that the stripe phase
was not present in our experiment.

Nevertheless, while $\aver Q$ matches the weak-coupling calculations for the
translationally-invariant B phase \cite{bib:VorontsovSauls:Aphase},
$\aver q$~is found to be reduced, see Fig.~\ref{fig:qQ}.
This contrasts with the good agreement between these calculations and
similar measurements in a $D=0.7$\,\um{} slab at higher pressure
\cite{bib:us:SlabPRL,bib:supplementary}, ruling out strong coupling effects
as the origin of this discrepancy.
We therefore consider domain structures in which the amount
$a$ and $1-a$ of domains with positive and negative $\Dperp$ is unequal \cite{bib:footnote:Dperp}.
For a qualitative estimate we assume that the domain walls are step-like
and ignore gap variation across the slab, $\Dpar = \const, |\Dperp|= \const$. Then
\begin{equation}\label{eq:crude:fraction}
\aver Q = \frac{|\Dperp|}{\Dpar}, \quad \aver q = (2a - 1)\frac{|\Dperp|}{\Dpar}, \quad \frac{\aver q}{\,\aver Q\,} = 2a - 1,
\end{equation}
demonstrating that while $\aver q$ is sensitive to the presence of domains,
to the first approximation $\aver Q$ is not, in agreement with our observations.
The gradual variation of $\Dpar$, $\Dperp$ and the emergence of the off-diagonal
gap matrix elements inside the domain walls would lead to corrections
to this model that should be taken into account in future theoretical work.

\begin{figure}[t!]
\includegraphics[width=0.48\textwidth]{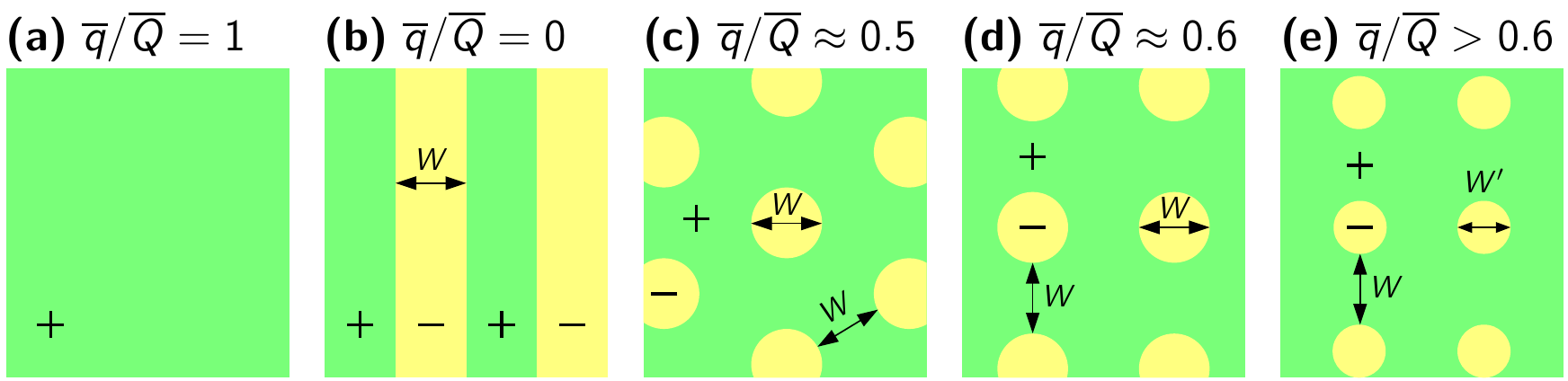}
\caption[Possible regular domain configurations in \He3-B confined to a slab]{Possible regular
domain configurations in \He3-B confined to a slab, view perpendicular to the slab plane.
$+/-$ indicates the sign of $\Dperp$.
(a) Translationally-invariant B phase, (b) predicted stripe phase,
(c,d) proposed polka dot phase that is easier to nucleate than the stripe phase.
The values of the gap distortion parameter $\aver q / \aver Q$ are derived under
a simplified assumption of step-like domain walls, Eq.~\eqref{eq:crude:fraction}.
Pinning of domain walls in the experimental cell may introduce disorder to these structures.
All features in (c-d) are taken to be of characteristic size similar
to the pitch $W$ of the stripe phase. (e) A variant of (d) in which dot diameter $W'$
is smaller than dot separation $W$.}
\label{fig:domainmorph}
\end{figure}

Our measurement $\aver q/\aver Q = 0.6 \pm 0.1$ near $T_{\text{AB}}$
suggests a $+/-$ domain proportion of $4:1$.
A likely scenario for an imbalance of the domains is a two-dimensional structure.
Within possible regular morphologies, Fig.~\ref{fig:domainmorph}, 
this imbalance corresponds to a \emph{polka dot phase} with hexagonal or square symmetry,
Fig.~\ref{fig:domainmorph}c,d.
Such structures have been suggested theoretically \cite{bib:VorontsovSauls:stripes},
but the detailed analysis of their energetic stability has not been carried out.
A preliminary Ginzburg-Landau study finds the square lattice of dots,
Fig.~\ref{fig:domainmorph}d, to be locally stable, with free energy
only slightly higher than that of the stripe phase~\cite{bib:Mizushima:dots}.
Further theoretical work is required to understand the stability of various
modulated states in the presence of strong coupling effects.

Even if less energetically favourable than stripes, the dots may arise because
of a lower energy barrier for flipping the sign of $\Dperp$ in a microscopic dot,
compared with a stripe, that is macroscopic in one dimension.
A lattice of dots would form if the neighbouring dots nucleate close
enough to prevent them from growing beyond a typical size $W$ before
getting within $W$ of the others.

Our experimental protocol is first to cool deep into the B phase,
in order to destabilise the domain walls, and then to take data on warming.
The key observation of the decrease in $\aver q/\aver Q$
with increasing temperature is consistent with the formation of negative-energy domain walls
in the B phase approaching the transition into the A phase, as predicted theoretically.
The observation of a single NMR line implies that the domain size is shorter than $\xi_{\text D}$.
The measured temperature dependence of $\aver q/\aver Q$ can be explained
by allowing the separation between dots $W$ and their diameter $W'$ be unequal,
see Fig.~\ref{fig:domainmorph}e.
Pinning of the domain walls by scratches on the cavity walls \cite{bib:pinning}
may play a role in restricting $W'$, and introduce disorder into the domain morphology.
Improved cavities have been developed for future experiments \cite{bib:siglass}.

As an alternative scenario, we now consider metastable domain walls with positive energy.
Defects are known to form at the A-B transition either due to inhomogeneous nucleation
\cite{bib:relic:wall,bib:Helsinki:BEC} or as relics
\cite{bib:Helsinki:HQVboundWalls} of defects, present in the A phase
at the start of the transition \cite{bib:Manchester:Achirality,bib:A:MRI}.
These may include the domain walls where $\Dperp$ changes sign \cite{bib:Thuneberg:hardwalls,
bib:Muh:halfQcirc,bib:Grenoble:vibwire,bib:Helsinki:HQVboundWalls},
which, if produced at unusually high density,
would result in a reduced $\aver q/\aver Q$ ratio;
however this ratio would remain constant if the defects are pinned
or increase with time as they decay, contrary to our observation.
This does not rule out sparse positive-energy defects with typical separation larger than 
$\xi_{\text D}$, giving rise to small satellite NMR signals, specific to each type
of defect \cite{bib:Helsinki:HQVboundWalls,bib:vortex:sheet,bib:half:quantum,bib:A:MRI}.
Detection of such signals is beyond the scope of this work.
Within errors our observations are independent of the rate of cooling through the A-B transition,
Fig.~\ref{fig:qQ}. This supports our proposal that defects produced at this transition
do not play a major role in the formation of domains on micron scale.
A systematic study of the influence of the cooling rate will be subject of future work.

In conclusion our NMR study of superfluid \He3 confined in a 1.1\,\um{} cavity in
the vicinity of the AB transition has found neither the predicted stripe phase,
nor translationally-invariant planar-distorted B phase.
This leads us to propose a superfluid phase with two-dimensional spatial modulation,
in a form of a regular or disordered array of island domains, driven by
negative energy of domain walls under confinement.
Further systematic studies of the nucleation of this phase, to determine the equilibrium
morphology, as well as its stability as a function of pressure,
predicted to be influenced by strong coupling effects, are both desirable.
Superfluid \He3 under confinement appears to provide a clean model system for spatially
modulated superconductivity/superfluidity, long sought in a wide variety of physical systems.

\begin{acknowledgments}
We thank B.~R.~Ilic for help with microfluidic chamber fabrication and design methodology;
A.~B.~Vorontsov and J.~A.~Sauls for sharing calculations
of the gap profile of confined $^3$He;
T.~Kawakami and T.~Mizushima for a stimulating discussion and for sharing
their preliminary result on the stability of the polka dot phase.
This work was supported by EPSRC grants EP/J022004/1 and EP/R04533X/1;
NSF grants DMR-1202991 and DMR-1708341, and the European Microkelvin Platform.
\end{acknowledgments}

\onecolumngrid

\vskip2.5em

\renewcommand{\thefigure}{S\arabic{figure}}
\renewcommand{\thetable}{S\arabic{table}}
\renewcommand{\theequation}{S\arabic{equation}}
\setcounter{figure}{0}
\setcounter{table}{0}
\setcounter{equation}{0}

\centerline{\large\textbf{Supplemental Material}}
\vskip1em

\twocolumngrid 

\begin{figure}[b!]
\includegraphics{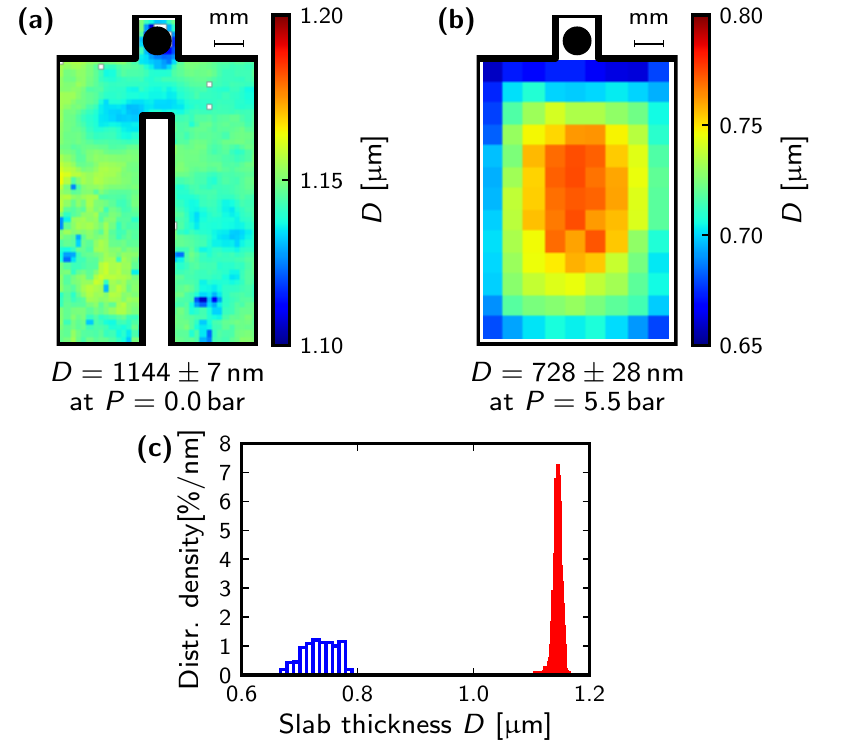}
\caption[Spectroscopic measurement of cavity thickness]{Spectroscopic
measurement of the thickness $D$ of silicon-glass microfluidic cavities.
The 1.1\,\um{} cavity used in the experiment described here is compared
to the 0.7\,\um{} cavity from Refs.~\cite{bib:us:SlabScience, bib:us:SlabPRL},
inflated under pressure necessary to stabilise the B phase.
In the present work the slab uniformity is improved, one of the reasons being
the partition wall in the middle of the cell \cite{bib:us:JLTP}.}
\label{fig:twocells}
\end{figure}

\section{Experimental Setup}
The apparatus of this experiment is in most aspects identical to that used
in Ref.~\cite{bib:us:SlabScience} and described in its supplementary material.
The main difference is the improved uniformity of the cavity,
see Fig.~\ref{fig:twocells}, achieved through optimised nanofabrication,
introduction of a partition wall and restricting the measurements
to low pressure~\cite{bib:us:JLTP}.
In addition the cell fill line was interrupted with a superfluid-leaktight
cryogenic valve~\cite{bib:us:JLTP,bib:coldvalve} at 6\,mK,
to avoid gradual depletion of the \He4 film due to the fountain effect.

\section{Temperature Correction}

\begin{figure}[b]
\includegraphics{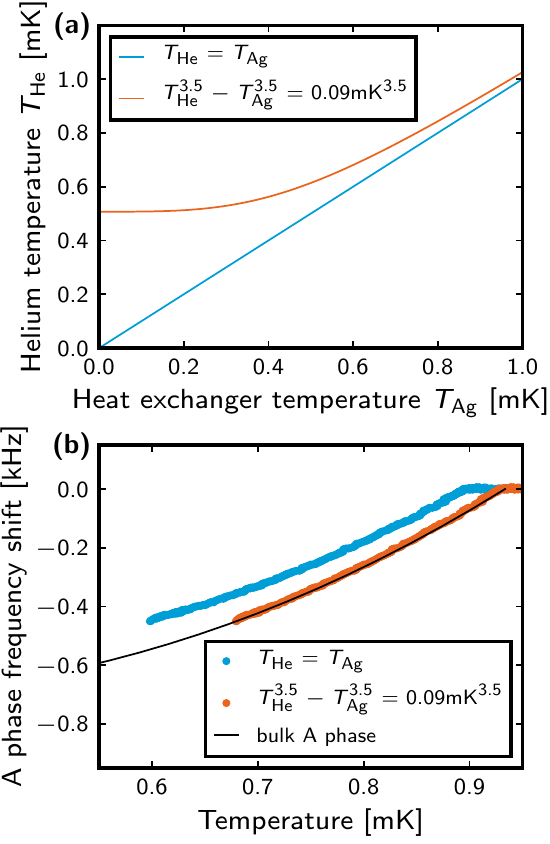}
\caption[Temperature correction]{Temperature correction. (a) Relationship between
temperatures $T_{\text{He}}$ and $T_{\text{Ag}}$ of helium and silver sinter in the heat
exchanger. (b) Frequency shift in the A phase is expected to have the bulk value,
since the energy gap in the A phase is not suppressed in a slab with
specular walls.
This is found to be the case after the temperature correction.
Here the bulk frequency shift is modelled based on the initial slope 
$2 f_{\text L} \partial\big|\Delta f(T)\big| / \partial(1 - T/T_c)|_{T \to T_c}
= 3.96 \times 10^{9}\,\text{Hz}^2$ measured in the $D=0.2$\,\um{} cavity with
98\%-specularly scattering walls and the calculated temperature dependence
of the weak-coupling A phase energy gap. Here the initial slope is defined as the slope
of a straight line fitted between $0.9 T_c$ and $1.0 T_c$.}\label{fig:Tcorr}
\end{figure}

\begin{figure*}[t!]
\includegraphics{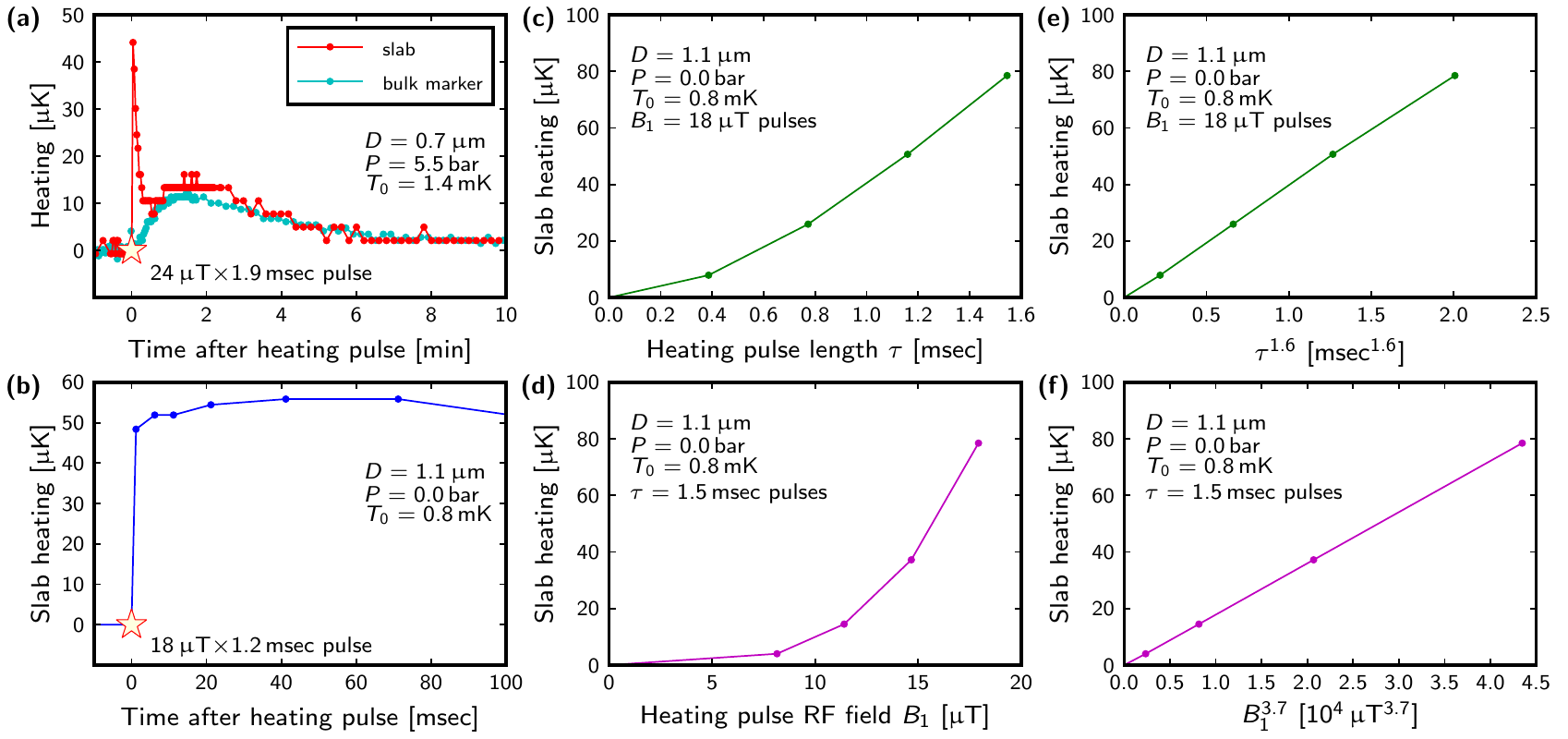}
\caption[Heating of the confined 3He by NMR pulses]{Characterisation
of heating of the confined \He3 by NMR pulses.
Helium temperature is inferred from frequency shift probed regularly
with small NMR pulses,
while the heating is induced by large pulses, applied sufficiently far
from Larmor frequency not to tip \He3 spins and rarely enough
(once every 20 minutes) to let the sample cool down before it is heated again.
(a) Heating in the slab and the ``bulk marker'' volume at the mouth
of the fill line. Only the former rapidly warms up due to local heating
we are concerned with; in addition both slowly respond to Joule heating
of the metallic elements of the experimental setup,
transmitted via the heat exchanger. Data from the $D=0.7$\,\um{} slab
with a large clearly-visible bulk marker \cite{bib:us:SlabScience}.
The rest of the measurements were performed on the $D=1.1$\,\um{} slab
presented in this paper.
(b) Time evolution of the slab temperature shortly after a large pulse.
(c-f) Dependence of heating on the pulse duration $\tau$ and
power of radio-frequency field $B_1$ measured 20\,msec after pulses, when the slab
reaches its highest temperature.}
\label{fig:heating}
\end{figure*}

The thermometry in our experimental setup is based on monitoring the temperature
of the silver sinter heat exchanger with a $^{195}$Pt NMR thermometer,
calibrated against the \He3 melting curve. 
In recent work on a $D=0.2$\,\um{} slab \cite{bib:us:200nm}
the temperature gradient between silver sinter (at $T_{\text{Ag}}$)
and helium (at $T_{\text{He}}$) in the heat exchanger was
carefully determined for various surface \He4 coverages.
With \He4 plating similar to the one used in this experiment,
the correction was found to be
\begin{equation}\label{eq:T35}
T_{\text{He}}^{3.5} - T_{\text{Ag}}^{3.5} = C,
\end{equation}
corresponding to thermal boundary resistance $R_{\text K}(T) \propto 1/T^{2.5}$.
The constant $C$ in \eqref{eq:T35}, determined by heat leak to the helium sample,
is obtained from the suppression of bulk $T_c$, registered with NMR
in the bulk \He3 marker at the mouth of the fill line.
Fig.~\ref{fig:Tcorr} demonstrates that such correction
with a slightly different $C$ is adequate for the present experiment,
which utilised the same heat exchanger.
In this work the NMR pulses were applied rarely enough to have no effect on $C$.

\section{Heating by NMR Pulses}

\begin{figure}[b!]
\includegraphics{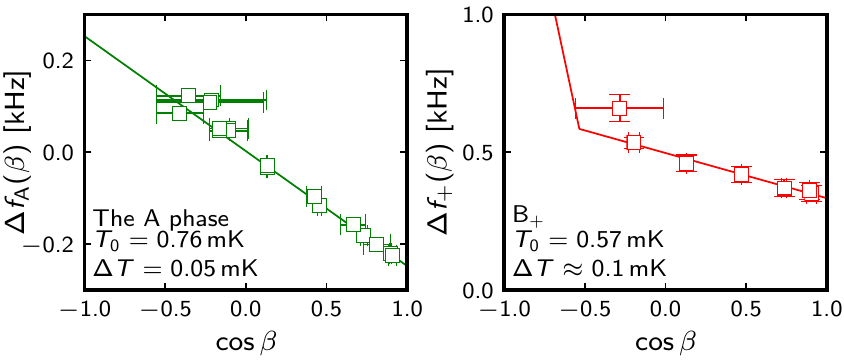}
\caption[Large tipping angle NMR measurements]{NMR measurements
on the A and B phase in $D=1.1$\,\um{} slab with 18\,\micro T$\times$1.2\,msec pulses.
These pulses overheated the slab by $\Delta T$ above the temperature $T_0$ of the
helium in the heat exchanger. Different tipping angles are achieved by applying the
initial part of the pulse with a 180$^\circ$ phase shift, see Fig.~2b.
In the dipole-unlocked A phase $\Delta f_{\text A}(\beta) \propto \cos\beta$,
see \cite{bib:us:SlabPRL} for $\Delta f_+(\beta)$.}\label{fig:NMR:large}%
\end{figure}

\begin{figure*}[t!]
\includegraphics{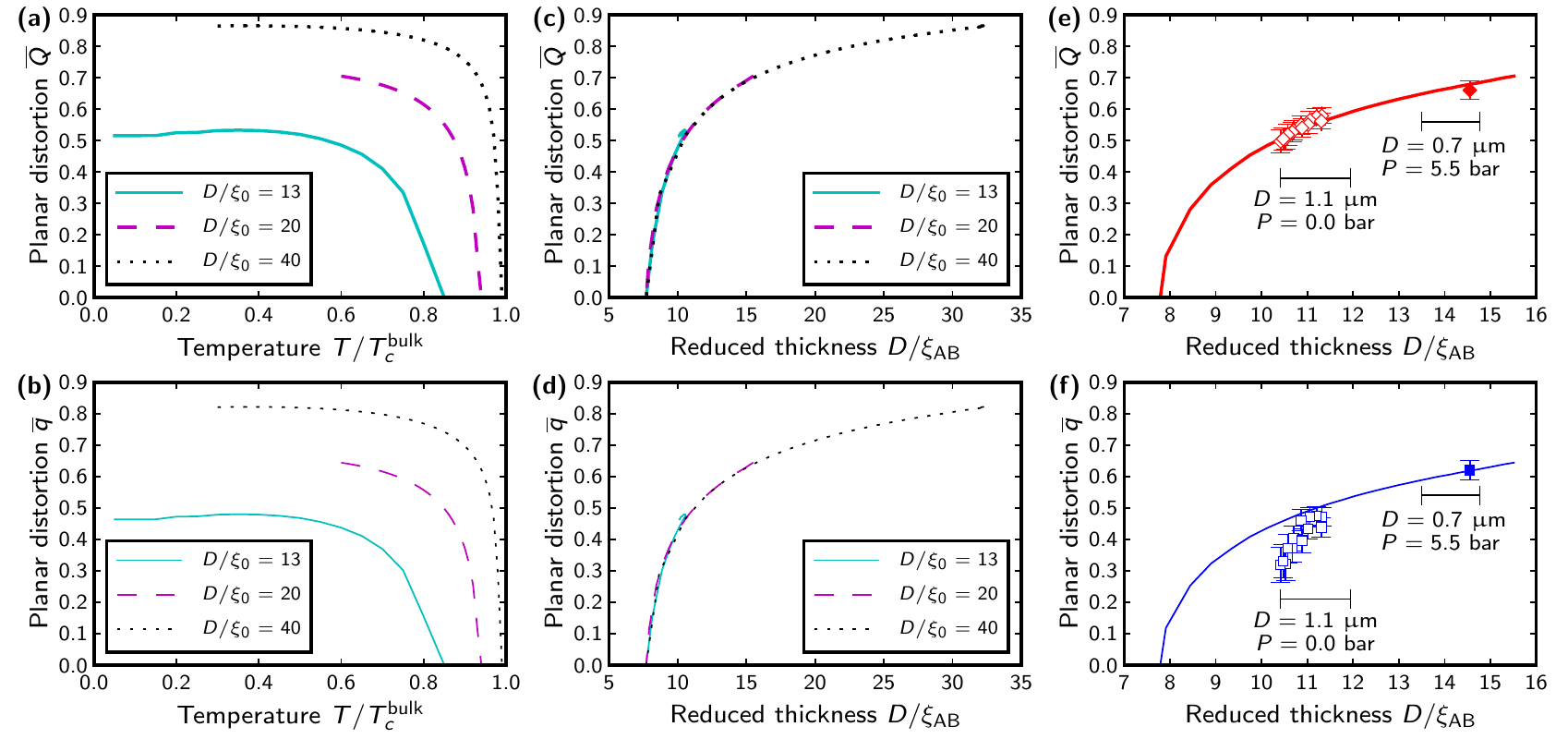}
\caption[Universal scaling of planar distortion]{Universal scaling of
planar distortion parameters $\aver q$ and $\aver Q$
with reduced slab thickness $D/\xi_{\text{AB}}$.
(a-d) Calculations of planar distortion of spatially-invariant planar-distorted B phase
at various $D/\xi_0$ \cite{bib:VorontsovSauls:Aphase}.
See Eq.~\eqref{eq:xi:AB} for the definition of the coherence length.
(e-f) Comparison of these calculations to measurements on 0.7\,\um{} slab
(filled symbols) \cite{bib:us:SlabPRL} and 1.1\,\um{} slab (open symbols,
showing together both warm-ups presented in Fig.~3).
The horizontal bars represent the range of $D/\xi_{\text{AB}}$
that can be probed in at given $D$ and $P$ limited by $T=T_{\text{AB}}$ and $T=0$.}
\label{fig:qQvsDxi}
\end{figure*}

NMR experiments on superfluid \He3 under regular confinement in silicon-glass
and fully-silicon cavities
(Refs.~\cite{bib:us:SlabScience,bib:us:SlabPRL,bib:us:200nm} and this work)
have manifested heating of an unidentified origin that couples to the confined liquid
directly, Fig.~\ref{fig:heating}a.
Most heating occurs within 1\,msec after the pulse
(the duration of the probe NMR pulse plus the dead time of the NMR spectrometer),
and the slab temperature is nearly constant for many msec afterwards,
Fig.~\ref{fig:heating}b.
The dependence of the heating on duration $\tau$ and amplitude $B_1$
of the sine-wave pulses, Fig.~\ref{fig:heating}c-f, is steeper than
expected for linear dissipation ($\dot Q \propto \tau B_1^2$),
especially when taking into account that specific heat of \He3 in the slab
increases with temperature. This points towards an exotic origin of this
parasitic effect.

In addition to the signatures shown in Fig.~\ref{fig:heating}
we found that the heating is not resonant near the \He3 Larmor frequency,
and that heating due to pulses with the initial `antipulse',
Fig.~2b, only depends on the total duration and amplitude,
but not on the length of the `antipulse' part.

As shown in Fig.~\ref{fig:heating}b, the free induction decay after
a large pulse with characteristic $T_2^* \approx 3$-10\,msec occurs
at a nearly constant temperature, elevated above the temperature $T_0$
of helium in the heat exchanger and fill line.
The use of `antipulses', Fig.~2b, allows us to measure the NMR response
as a function of tipping angle at a \emph{constant} elevated temperature.
This is illustrated in Fig.~\ref{fig:NMR:large}. We note that the strong frequency
dependence of NMR tipping by `antipulses',
renders them unusable away from their carrier frequency.
For this reason B$_-$ is missing from Fig.~\ref{fig:NMR:large},
due to relatively large frequency range spanned by
$\Delta f_-(\beta) \approx -1\,\text{kHz} \times \cos(\beta)$.
Probing B$_+$ above the magic angle $\beta^*$ is equally
problematic and was not studied here in detail.

In contrast the 0.7\,\um{} slab was probed with groups of large pulses that caused
different heating, restricting such measurements to the $T \lesssim 0.5 T_c$ limit
where the temperature dependence of the frequency shifts is weak~\cite{bib:us:SlabPRL}.

In this work we demonstrated that tipping angles up to $60^\circ$ could be reached with
8\,\micro{}T$\times$1.2\,msec pulses, generating only $\Delta T \lesssim 10$\,\micro K,
small compared to the temperature range over which we probed the B phase gap distortion.

\section{Universal Scaling of Planar Distortion}

Within the Ginzburg-Landau regime, $T - T_c \ll T_c$, the effects of
confinement on properties of the superfluid are determined by a single
control parameter, the reduced thickness $D/\xi(T,P)$, where
$\xi$ is the coherence length.
This universality breaks down at lower temperatures, i.e.~see
the supplementary of Ref.~\cite{bib:us:SlabScience}.
To study the A-B transition outside of the Ginzburg-Landau regime
the coherence length has been defined as
\begin{equation}\label{eq:xi:Delta}
\xi_\Delta(T,P) = \frac{\hbar v_{\text F}(P)}{\Delta_{\text B}(T,P) \sqrt{10}},
\end{equation}
were $v_{\text F}$ is the Fermi velocity and $\Delta_{\text B}$ is the
bulk B phase gap. We observe that the weak-coupling quasiclassical
calculations~\cite{bib:VorontsovSauls:Aphase} of $\aver q(T)$ and $\aver Q(T)$
at different $D$ collapse, see Fig.~\ref{fig:qQvsDxi}a-d, expressed
as a function of $D/\xi$ adopting a slightly different coherence length 
\begin{equation}\label{eq:xi:AB}
\xi_{\text{AB}}(T,P) = \frac{D_{\text{AB}}(T,P)}{\pi\sqrt{6}},
\end{equation}
where $D_{\text{AB}}(T,P)$ is the thickness of the slab at which
the A to B transition occurs at temperature $T$ and pressure $P$
(the inverse of the $T_{\text{AB}}(D/\xi_0(P))$ function \cite{bib:Nagai:AB}).
Here we restrict the discussion to the calculations for a slab
with specular boundaries and recognise that the pressure $P$
only enters the weak-coupling calculations via the pressure dependence of
the bulk transition temperature $T_c^{\text{bulk}}(P)$ and the Cooper pair diameter
$\xi_0(P) = \hbar v_{\text F}(P) / 2\pi k_{\text B} T_c^{\text{bulk}}(P)$.
We find that $\xi_{\text{AB}} / \xi_\Delta \to 1.0$ at $T\to T_c$
and $\xi_{\text{AB}} / \xi_\Delta \to 1.1$ at $T\to 0$.

\section{Planar Distortion: Theory vs Experiments}

We compare the calculations discussed in the previous section
with the NMR measurements of the planar distortion of the B phase in 0.7 and 1.1\,\um{} slabs
in Fig.~\ref{fig:qQvsDxi}e,f.
The former, obtained at $P=5.5$\,bar and $T=0.6\,\text{mK}=0.4 T_c^{\text{bulk}}$ are in good
agreement with the theory in terms of both $\aver q$ and $\aver Q$.
This confirms that the strong coupling effects, known to increase with pressure,
are not responsible for the reduced $\aver q$ presented in this paper, Fig.~3
and Fig.~\ref{fig:qQvsDxi}f.

\section{Qualitative Discussion of Domain Configurations}

In this section we compare possible structures of domains using
the length of domain walls per unit area of the slab $L/A$ as a
figure of merit of the free energy gain due to formation of domain walls.
We consider a two-dimensional problem of energetic stability
of thin domain walls with hard-core repulsion at distance $W$.
Dots are assumed to be circular.
We find $L/A = 1/W$ for the stripe phase, $L/A = \pi/2 W\sqrt{3} \approx 0.91 / W$
for hexagonal lattice and $L/A = \pi/4 W \approx 0.79 / W$ for square lattice,
listed in Fig.~4b-d. The fact that these numbers are all close to each other demonstrates
that the free energy of 1D and 2D modulated states is nearly degenerate, and only
a detailed calculation can reliably identify the lowest energy state,
taking into account the gradual spatial variation of the order parameter components
across the domain walls and the optimum shape of dots.

\rule{0pt}{24em}

\rule{0pt}{61em}


\begin{thebibliography}{99}
\bibitem{bib:stripe:sign} Note that the state with $\Dperp < 0$ on the right of the domain wall
in Fig.~\ref{fig:domainwall} can be written in the canonical form Eq.~\eqref{eq:Bplanar}
with $\Dperp > 0$ using a $\pi$ phase shift and a combined rotation
$\mtrx R(\hvec n', \theta') = \mtrx R(\hvec n, \theta) \mtrx R(\hvec z, \pi)$:
\begin{multline*}
e^{i\phi} \mtrx R(\hvec n, \theta) \begin{pmatrix}\Dpar & &\\ & \!\!\Dpar\!\! & \\ & & \!\!\!\!-\Dperp\end{pmatrix} =
-e^{i\phi} \mtrx R(\hvec n, \theta) \begin{pmatrix}-\Dpar\!\! & &\\ & \!\!\!\!-\Dpar\!\! & \\ & & \Dperp\end{pmatrix} \\
 = e^{i(\phi + \pi)} \mtrx R(\hvec n, \theta) \mtrx R(\hvec z, \pi) \begin{pmatrix}\Dpar & &\\ & \!\!\Dpar\!\! & \\ & & \Dperp\end{pmatrix}.
\end{multline*}
\bibitem{bib:footnote:Dperp}
Since the sign of $\Dperp$ can be flipped throughout the sample with a uniform transformation \cite{bib:stripe:sign},
we only need to consider the case of $\Dperp$ being positive in the larger part of the slab,
$a \ge 0.5$, so $\aver q \ge 0$, as required in Eq.~\eqref{eq:fshiftqQ} \cite{bib:us:SlabPRL}.
\end{thebibliography}
\end{document}